\documentclass{article}
\begin{document}

\begin{center}
\centerline{\large \bf Coherency is the ether of XXI century.}
\end{center}

\vspace{3 pt}
\centerline{\sl V.A.Kuz'menko}
\vspace{5 pt}
\centerline{\small \it Troitsk Institute for Innovation and Fusion 
Research,}
\centerline{\small \it Troitsk, Moscow region, 142190, Russian 
Federation.}
\vspace{5 pt}
\begin{abstract}

	The concept of coherent states in explanation of a nature of nonlinear 
phenomena in optics will be inevitably replaced by the concept of inequality 
of forward and reversed transitions.

\vspace{5 pt}
{PACS number: 42.50Gy, 42.50.Hz}
\end{abstract}

\vspace{12 pt}

The concept of ether exists for many centuries. But in the twentieth century 
its necessity practically comes to the end and for most physicists this idea 
remains only as a historical curious thing.

	In substantial degree similar future, obviously, wait for the concept 
of coherency, which now widely used for explanation of nonlinear phenomena in 
optics. Formerly such effects usually are explained as a nonlinear response 
of substance on high intensity laser radiation [1]. In fact this is not an 
explanation, but only a certification of result: if the substance has high 
nonlinear optical susceptibility coefficient of second, third or higher order, 
then the high intensity laser radiation will give corresponding nonlinear 
effect. The Maxwell-Bloch equations or the rotating wave approximation is 
usually used for mathematical description of nonlinear phenomena. This 
mathematical model in many cases gives good description of nonlinear effects, 
but it does not have clear physical sense and the physical origin of nonlinear 
phenomena should be explained independently. 

	The concept of coherency now plays a main role in explanation of 
origin of laser-driven nonlinear optical effects. It is supposed, that laser 
radiation creates a so-called coherent states. Those states have specific 
properties and can interfere. As a result the nonlinear phenomena appear, 
like as coherent population trapping [2], electromagnetically induced 
transparency [3], amplification without inversion [4,5], etc. Although it is 
not sufficiently clear what is the nature of coherent state and how its can 
interfere, such explanation looks likely. It is assumed that collisions 
must destroy coherent states. And really the experimentalists observe, 
that collisions destroy the discussed nonlinear effects. As a result, 
the concept of coherency become  all pervasive in the field of laser and 
quantum optics [6]. 

	However, a skilled theoretical analysis had discovered the inward 
defect in the coherency concept [7,8]. It has been shown that the inability 
to measure the absolute phase of an electromagnetic field prohibits the 
existence of coherent states. In fact it means that the using of concept of 
coherent states for explanation of nonlinear phenomena does not have physical 
sense. This result is a shock for experimentalists. It robs and nothing 
gives in exchange for their favorite toy. The experimentalists can not 
agree with such situation. They exert strong pressure on theorists. As a 
result, some attempts to revise this theoretical conclusion appear. The work 
[6] contains excellent description of the situation as a whole and makes 
attempt to save the concept of coherent states by searching the weak point 
in arguments of works [7,8]. 

	The goal of this note is not to discuss the arguments. The goal is 
to point out that we do not need to save the concept of coherent state. 
There exists alternative concept, which better suit for explanation of 
nature of nonlinear effects. We keep in mind the concept of inequality of 
forward and reversed optical transitions, which corresponds to concept of 
time invariance violation in electromagnetic interactions [9]. 

	The orthodox point of view claims, that the time invariance violation 
is absent in electromagnetic interactions. This point of view does not have 
experimental proofs, but it has a long historical tradition [10]. In contrast, 
the experimental proofs for the opposite point of view appear recently. 
First of all this is an experimental study of forward and reversed 
transitions in optics. Seemingly, it is very simple to test the invariance 
of a photon absorption process: we should measure the parameters of photon 
absorption process (spectral width and cross-section) and compare its with 
the measured parameters of the reversed stimulated emission process. But 
really the situation is not so simple. Both processes take place 
simultaneously. Furthermore, the spectral width of laser radiation or optical 
transition are usually connected accordingly with the pulse length and life 
time of excited states toward a spontaneous emission. All this reasons make so 
entangled situation, that reliable measurement and comparison of the 
necessary parameters of forward and reversed processes become perfectly 
problematic. 

	However, in one unique for present day case this problem is easily 
overcame. This is the case of the so-called line wings in the absorption 
spectrum of polyatomic molecules [9]. This physical object has unusual  
properties: long lifetime of excited states toward a spontaneous emission 
is combined with extremely large homogeneous spectral width of optical 
transition. In this case simple pump-probe experiments in a molecular beam 
conditions clearly show, that the spectral width for the reversed transition 
is in several order of magnitude smaller, than those for the forward 
transition, accordingly, the cross-section of the reversed transition is in 
several order of magnitude grater, than the cross-section of the forward 
transition [11]. 

	For present day there is the first and the most clear and reliable 
experimental proof of time invariance violation in optics. Other experimental 
proof was obtained recently in the study of interaction between light and 
metallic planar chiral nanostructures [12,13]. Besides that, the quite formal 
fresh sight analysis of high order nonlinear phenomena in optics shows theirs 
violation of time reversal symmetry [14]. 

	Time invariance violation is very good foundation for explanation of 
physical origin of nonlinear effects. Inequality of forward and reversed 
transitions endows atoms and molecules the properties of memory about the 
initial state and aspiration to return in this state. This properties look 
like very similar to those properties of coherent states: the memory also 
must be destroyed by collisions. This properties allow to give simple and 
clear explanation of physical origin of most nonlinear phenomena [9,15]. 

	So, we believe, that the concept of inequality of forward and 
reversed transitions will inevitably replace in nonlinear optics the 
concept of coherent states. Because of:

1) it much better explain the nature of most nonlinear phenomena,

2) in contrast to the concept of coherent states, it has a real physical 
base [16,17], 

3) in contrast to the concept of coherent states, it has direct experimental 
proofs [9,12,13]. 

There is possible other way also. Because of the myths (like as the ether 
or the coherent states) have not usually a strict definition, we can preserve 
the brand of the concept, but substitute for it contents. Coherent states can 
be defined as a specific states of atoms and molecules after forward optical 
transitions. However, such variant may be inconvenient because of it makes 
some mishmash for readers (such situation exists now in rather similar case 
with the concept of the "quasicontinuum" of vibration states of polyatomic 
molecules [18]). 

In conclusion, the orthodox point of view is very strong now. The 
experimentalists are afraid to work in the new direction. They need to be 
aimed by the theorists. But the theorists use their efforts now for saving 
the hopeless idea. So, dear colleagues, look around and let us begin to work 
in the right direction at last.

\end{document}